\documentstyle[epsffrag,rotate,graphicx,psfrag]{elsart}
\begin{document}
\begin{frontmatter}
%
\title{On the theory of vortex quantum tunnelling in the dense Bose
superfluid helium II}
\author[address1]{Uwe R. Fischer\thanksref{thank1}}
\address[address1]{Low Temperature Laboratory, 
Helsinki University of Technology, 
P.O. Box 2200, FIN-02015 HUT}
\thanks[thank1]{Permanent address: Institut f\"ur Theoretische Astrophysik,
Universit\"at T\"ubingen, 
Auf der Morgenstelle 10, D-72076 T\"ubingen. E-mail: fischer@tat.physik.uni-tuebingen.de}  
\begin{keyword}
Quantum tunnelling, Vortices, Dense Superfluid
\end{keyword}
\begin{abstract}
The quantum tunnelling and nucleation theory of vortices 
in helium II is reviewed. 
Arguments are given that 
the only reliable method
to calculate tunnelling probabilities in this highly correlated,
strongly interacting many-body system is the semiclassical, large scale 
approach for evaluation of the tunnelling exponent,  
which does not make any assumptions about the unknown dynamical
behaviour of the fluid on microscopic scales. The geometric
implications of this semiclassical theory are represented in some
detail and its relevance for the interpretation of experimental
data is discussed.
\end{abstract}
\end{frontmatter}
\newcommand{\MF}{{\large{\manual META}\-{\manual FONT}}}
\newcommand{\manual}{rm}        
\newcommand{\bm}[1]{\mbox{\boldmath$#1$}}
\newcommand{\kbar}{{{\bf --}}\hspace{-5.5pt}$\kappa$}
\maketitle
\section{Preface}
The nucleation theory of quantized vortices in the Bose superfluid 
helium II has been an elusive subject ever since the existence of quantized
vortices was conjectured by Lars Onsager in 1949. 
The difficulty to grasp their coming into existence  in a definite 
manner from first principles has one fundamental reason: 
There is no microscopic theory of this superfluid. 
We do not know how to describe the motion of a vortex on scales 
of the order of the coherence length, where this motion is governed 
by the full quantum many-body structure of the superfluid and the interaction 
of the vortex with the microscopic excitations. 
Consequently, we can not follow the vortex 
on any stage of
the evolution from a virtual vortex to a mature one describable by 
semiclassical means. 
At present, even the many-body structure and energy 
of a simple static rectilinear vortex can not be uniquely 
determined, let
alone that of a circular vortex which is still beyond reach even for 
Monte Carlo calculations \cite{ortiz}.

\begin{center}
\begin{figure}[hbt]\label{scales}
\rotate[r]{\includegraphics[width=0.6\textwidth]{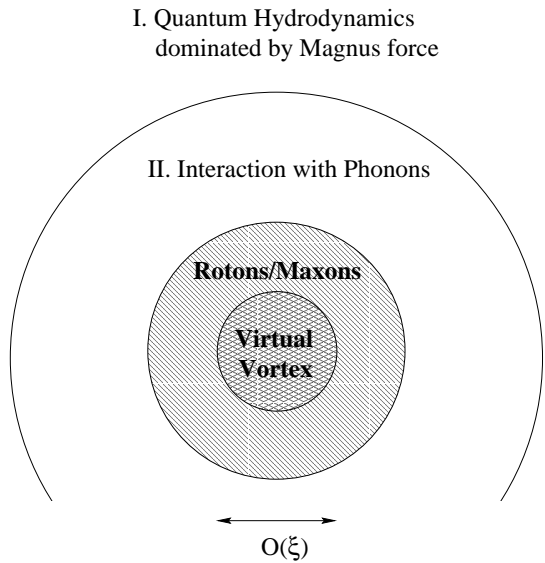}}
\caption{Length scales of vortex motion and interaction with elementary 
excitations in the dense  superfluid helium II. It is only 
in the regions I and II where a semiclassical 
picture of vortex motion is applicable. The vortex in the central
region of extension O($\xi$) is termed virtual because it is no
well-defined topological object in this region.}
\end{figure}
\end{center}

This lack of a microscopic idea of vortex motion makes it necessary to
resort to a 
semiclassical, hydrodynamic (large scale) theory. In such a treatment,
the existence of the vortex as a semiclassical object
has to be assumed {\it ab initio}. No details of
the underlying microscopic dynamics, {\it i.e.} of the actual 
nucleation event, are to be described in such a theory, 
but only the laws which rule vortex motion on curvature scales well beyond 
the core size $\xi$ (Fig. \ref{scales}).  
This is admissible because in helium II
this length scale is of the order of the 
interatomic spacing and the relevant scales for potential barriers to be 
surmounted by thermal activation or crossed by 
quantum mechanical tunnelling
usually have spatial extent and length scales far beyond $\xi$. 
Other advantages of $^4\!$He 
include that it can be prepared to have very high purity such that dissipation
has negligible influence. 
The order parameter $\phi$ 
has the simplest conceivable structure resulting from a spontaneously
broken U(1) symmetry.
In contrast to the extraordinarily complicated behaviour on $\xi$-scales
this superfluid can be studied under quite basic and simple 
assumptions if one is remaining completely in the
semiclassical, large scale domain. 

In this paper, we will discuss and investigate more closely 
the hydrodynamic theory of vortex quantum tunnelling.  
In the section which follows we contrast it with different other
ways to calculate tunnelling 
rates. 
Then we explain the way in which the
classical theory of the tunnelling exponent in three dimensions 
is a geometric theory in section \ref{semigeo}.
How to obtain an explicit solution of the boundary problem for a point vortex 
moving around an ellipse in two dimensions 
is demonstrated in section \ref{geodepend}.
It is shown that at boundaries with strongly varying curvature radii  
strict bounds on semiclassicality of vortex motion are required, which
lead in turn to restrictions on the tunnelling energy of the vortex. 
We estimate the prefactor of the tunnelling rate 
in section \ref{prefactor} and the relation to available
experimental data is established. We conclude with some general remarks.  
The discussion will be restricted to zero 
temperature as we will be interested in pure quantum mechanical 
tunnelling without any assistance of thermal fluctuations. 
\section{Different approaches}
We present
in this section a critical overview of some of the 
methods developed over the last decades to describe vortex quantum
nucleation and tunnelling.

The description of the tunnelling phenomenon by means of classical
incompressible hydrodynamics for the vortex motion was first 
undertaken in \cite{volovik}. The motion of a vortex half-ring in the presence 
of a half-sphere at an otherwise flat boundary 
was considered, which is solvable analytically  because of maximal symmetry
by using a Legendre function of the second kind. The method is exact 
insofar as the (pinning) potential in which the vortex moves is exactly known
in its relation to known geometrical quantities. 
This procedure  involves no assumptions 
about the dynamics of the order parameter. The only fundamental ingredient
is the existence of the Magnus force acting on the vortex. 
This is essentially the approach which we will investigate in this
paper. 

The approach of \cite{sonin} consisted in evaluating transition 
probabilities between an initial many-particle wave function without
and a final one with vortex. The many-particle wave functions were 
constructed from single particle solutions of the Gross-Pitaevski\v{\i}
equation \cite{GP}, 
valid for a weakly interacting, dilute Bose gas. The main difference
to \cite{volovik}, as discussed in \cite{sonin3}, consisted in the existence 
of an (intermediate) deformed vortex ring state with a depression 
of the flow velocity out of the ring plane, leading to a smaller
vortex energy. This also led to a different 
form  of the tunnelling exponent by a logarithmic factor.  
Ultimately, though, this theory gave approximately the same value
for the exponent  as the 
collective co-ordinate approach of \cite{volovik}, provided one takes 
values of tunnelling parameters typically realized. 

In the recent past, there has been put some effort into 
numerical calculations of instantons 
\cite{freire,chudnovsky}.  
The instanton is a 
finite action solution 
to the Euclidean equations of motion and describes imaginary
time motion under the tunnelling barrier.  
In \cite{chudnovsky}, an effective Lagrangian for a massive elastic 
string moving in a pinning potential near criticality was derived
to calculate the shape of the instanton.   
The authors of \cite{freire}, making use of the 
Gross-Pitaevski\v{\i} 
equation, examined the nucleation of vortices 
in the streaming motion past an obstacle by solving 
the full nonlinear field equation in 1+1d. The intent
there was to investigate the dynamics of the vortex linked to that of
the condensate.   
With regard to this approach, it should be emphasised that   
use of the Gross-Pitaevski\v{\i} 
equation for helium II on small ($\xi$) scales can at best claim
phenomenological correctness, in this way comparable to 
density-phase functional theories \cite{dalfovo2}. 
  

Field-theoretical calculations, 
akin to those of pair
creation in quantum electrodynamics, but for  the corresponding 
one-dimensionally extended, closed  object vortex ring in a U(1) 
superfluid,  
were performed in \cite{teitelboim,davis3,davis5,kaolee}. 
The antisymmetric gauge field tensor
in the dual formulation of string dynamics \cite{kalb} 
has been employed and the components of the 
3-form field strength tensor identified with
stringy generalizations of magnetic and electric fields. The theory is
`relativistic' in the sense that the propagation velocity of light is 
replaced by that of sound in Lorentz invariant wave equations.         
The main problem of this formulation 
can be ascribed to the usage of 
an effective Nambu-Goldstone Lagrangian for sound on 
scales of O$(\xi)$.
This Lagrangian does not describe the real superfluid 
on these scales.
In particular, because vortex velocities approach
the speed of sound $c_s$, these approaches give the {\em
hydrodynamic}, {\it i.e.} large scale,  
mass of the vortex undue weight in the tunnelling exponent 
\cite{grisha2,stephen}. 

In summary, all these approaches have in common that they 
are {\em mean-field} theories, assuming that the quantum mechanical 
fluctuations of the U(1) field $\hat \phi ({\vec x},t)$ 
under consideration are on any scale relatively
small compared to its expectation value $\phi=<\hat \phi ({\vec x},t)>$, 
the order parameter.
Such theories can not be derived in $^4\!$He 
from first principles on arbitrary length scales, 
because this dense superfluid is a strongly coupled system, in which 
quantum fluctuations play a major role. On the atomic 
scale of the coherence length this implies, for example, 
that the quantum mechanical uncertainty for the density 
operator is of the same order as the local value of the density itself, 
the expectation value of this operator. Put very
simple: One does not know (for purely quantum statistical reasons)
if the particle(s) in a coherence length 
sized volume is (are) there or not. The variable conjugate to 
density, the phase, has no definite value in such a small volume as well.  
In other words still, a collective central 
vortex co-ordinate of the quantum field ceases to be meaningful for
the description of vortex motion on these small scales.   
\section{Semiclassical vortex quantum tunnelling as a geometric theory}
\label{semigeo}
The aforementioned reservations about various (mainly field-theoretical) 
approaches led the present author to a re-investi\-gation \cite{geo} of 
the method \cite{volovik} put forward already in 1972. 
The most appealing feature of this method is
its simplicity and exactness in the semiclassical limit 
(cf. regions I and II in Fig. \ref{scales}). 
There is no reason for the description of the 
vortex as a stringlike entity  in terms of a collective co-ordinate 
not to be a viable candidate       
for the investigation of vortex quantum tunnelling,  
if the requirements for this formulation are carefully met.
It should have become apparent from the above discussion 
that this method is in fact at present the only fully reliable
one for the calculation of observable tunnelling probabilities 
in the actual, dense superfluid helium II.   
\subsection{The vortex action for a string}
To set up the general framework for the description of a vortex string
in the hydrodynamic limit,  
it is convenient to use the dual formulation of string
dynamics in terms of the  antisymmetric tensor gauge field $b_{\mu\nu}$ 
already mentioned above in the field-theore\-tical context
\cite{myself}. 
The appendix \ref{Adual} contains a short summary of this  formulation.
The simplest way to get an intuitive picture of the dual formulation
is that the tensor gauge field is a relativistic generalization of the stream
function of classical hydrodynamics \cite{milne-t}, {\it i.e.} it is
chosen such that its (generalized) rotation is the velocity 
field of the fluid. 
To begin with and introduce further into this formalism as far as we need it
in what follows, we then recall in addition that a
semiclassical point vortex is analogous to a charged particle in a magnetic
field.
The stream function $\psi$ of classical hydrodynamics
plays the part of the negative scalar potential $-a^0\equiv-\Phi$ 
acting on the vortex
with a charge $q=(m\rho_0)^{1/2}\kappa$. The  
vector potential $a_i$ stems from the rotation of a `magnetic' field which is
antiparallel to the local circulation vector and  
related to the superfluid density.     
These {\it analoga} can be readily generalized to a stringlike entity by
defining the above quantities locally on the string \cite{myself}.  
The Magnus force, a stringy generalization of the Lorentz force,
is then derived in a Galilei invariant  superfluid from the
variation of the action 
\begin{eqnarray}\label{Magnus}
S_M & = &  
m \rho_0 \kappa \int\!\!\oint dt\, d\sigma 
\left(\psi_i X'^i + b_{ij}X'^j \dot X^i \right)\\ 
& \equiv & q\int\!\!\oint dt\, d\sigma 
\left(-\Phi  + a_i \dot X^i \right)\nonumber
\end{eqnarray}
with respect to the vortex co-ordinate $X^i (t,\sigma )$ (we employ the
convention of summation over equal indices). The line tangent 
is denoted $X'^i\equiv \partial X^i /\partial
\sigma $ and the vortex velocity $\dot X^i\equiv \partial X^i
/\partial t $. 
The bulk number
density and the quantum of circulation bear the symbols $\rho_0$ and
$\kappa$, respectively. The parameter $\sigma$ labels points on the
string.

As expressed in the second line of equation
(\ref{Magnus}),   
the first part of the action $S_M$ is the generalization of the negative
scalar potential in usual (non-relativistic) electrodynamic actions.
The second part is 
analogous to the linear coupling term of the gauge vector potential 
$a_i= (m\rho_0)^{1/2}b_{ij}X'^j$ to the vortex velocity $\dot X^i$. 
The generalization from the point particle to the string essentially 
consists in
introducing locality in $\sigma$ and admitting that the line tangent 
can point in any direction. The stream function becomes a vector 
with components $\psi_i=b_{0i}$ along the tangent vector. 
An isotropic gauge choice for the purely spatial part of $b_{\mu\nu}$
in Cartesian co-ordinates is given by $b_{ij}=-(1/d)n_{ijk}X^k$, where
$d$ signifies the spatial dimension and $n_{ijk}$ the unit
antisymmetric symbol. This represents the isotropic solution of equation
(\ref{bijk}).       

The local canonical 
momentum of a vortex per $\sigma$-length 
in a {\em compressible} superfluid 
\begin{equation}
P_i =  P^{\rm inc}_i + P^{\rm kin}_i=
m\rho_0\kappa\, b_{ij} X'^j + \sqrt \gamma\, M_0 \dot X_i \,
\end{equation}  
is a gauge dependent quantity. 
The quantity $\sqrt \gamma \, d\sigma$ is the measure of the string's proper
length. 
The vortex momentum consists 
of two terms, of which the gauge dependent first one, $P_i^{\rm inc}$,
is the contribution of the vortex-velocity dependent term in the 
Magnus action $S_M$ (it corresponds to the electrodynamical minimal 
coupling term $q a_i$). 
 The second one, 
the kinetic (vortex matter) part $P_i^{\rm kin}=\sqrt \gamma \,M_0
\dot X_i$, comes
into play if the  finite compressibility of the superfluid is taken
into account.  This hydrodynamic contribution to the vortex momentum is 
acquired from the self interaction of the vortex string with phonons.
The static mass of the vortex is in the present non-charged case 
$M_0= E_{\rm self}/c_s^2$, {\it i.e.} 
its static self energy $E_{\rm self}$ 
divided by the speed of sound squa\-red \cite{duan}.
In the given form the static mass is 
of relevance in the actual dense superfluid 
if vortex velocities remain semiclassical, that is, 
much less than the Landau critical velocity related to the quantum 
many-body structure of the superfluid.  
The hydrodynamic mass is, up to a factor of order
unity, given by $M_0\sim M_{\rm core}\ln[R/(\xi e^{C_0})]$, 
where $R$ is the infrared cut-off (distance to the next vortex or 
size of the vessel) and $C_0$ is a core constant, reflecting our
ignorance of the core structure.  
The core mass $M_{\rm core}=m\rho_0 \pi \xi^2 $ 
is a measure of the mass per unit length of the normal 
fluid contained in the core.
 
The contribution of ${\bm P}^{\rm kin}$
(coming mainly from region II in Fig. \ref{scales})  
is very small 
compared to that of the Magnus force acting in region I and II
\cite{grisha2,stephen,myself,geo}.
The reason lies in the fact that the ratio of these two quantities, 
\begin{equation}
\frac{\,\left|{\bm P}^{\rm kin}\right|\,}
{\,\left|{\bm P}^{\rm inc}\right|\,}
\approx \frac\kappa {|{\bm X}|c_s}\frac{|\dot {\bm X}|}{c_s}\,, 
\end{equation} depends on 
$|\dot {\bm X}|/c_s$ as well as $\kappa/(c_s|{\bm X}|) $ 
($= {\rm O}(\xi/|{\bm X}|)$ in helium II). 
Both quantities are necessarily $\ll 1$ if the vortex is to be
described semiclassically. We used here that $|b_{ij}|$ is of 
order $|X^k|$ in a Cartesian frame, and neglected  
the dependence of the self energy logarithm on the vortex co-ordinates.
 
The vortex action has the canonical form  
\begin{equation}\label{action}
S_V = \int_0^T\!\!\oint\! dt \, d\sigma \,
\dot {\bm X}\cdot {\bm P} - \int_0^T\! dt\,  H_V\, ,
\end{equation}
where the Hamiltonian $H_V$ is the sum of static, kinetic, elastic 
and potential terms \cite{myself}:
\begin{eqnarray}
H_V 
=  \oint\! d\sigma 
\sqrt\gamma\, \left[M_0 c_s^2  
+ \frac 1{2\gamma M_0} \left( {\bm P}
-{\bm P}^{\rm inc}\right)^2\right.\nonumber \\
 \quad \qquad \left. +\frac{M_0 c_s^2}{2\gamma}\,
{\bm Q}'{}^2\right] - m \rho_0 \kappa\int\! d\sigma \left(\frac12 {\bm \psi}_C 
 +{\bm \psi}_{u}\right)\cdot  {\bm X}'\,. \label{H}
\end{eqnarray}
Here, the vector $\bm Q$ is lying in the plane locally perpendicular to $\bm
X'$ and measures deviations from an equilibrium shape of the string.  
The stream function vector $\bm \psi$ (see Appendix \ref{Adual})  
is separated into its 
Coulomb part ${\bm \psi}_C$ from the interaction of the vortex in
question with other vortices and the part ${\bm \psi}_{u}$ stemming 
from the interaction with a (constant) background flow field. 
The first term in the square brackets represents the static, the 
second the kinetic, and the last one the elastic energy of
the string.
In this representation for $H_V$, we imposed the Coulomb gauge
condition ${\rm div} {\bm P}^{\rm inc}=0$, which results 
in the factor of 1/2 in front of the `Coulomb' potential 
${\bm \psi}_C$.
\begin{center}
\begin{figure}[hbt]
\psfrag{a}{\small $\alpha(\sigma)$}
\psfrag{sigma}{$\sigma$}
\psfrag{kappa}{$\kappa$}
\psfrag{Q}{${\bm P}^{\rm inc}$}  
\psfrag{Q1}{$P^{\rm inc}_{\it 1}$}
\psfrag{Q2}{$P^{\rm inc}_{\it 2}$}
\psfrag{X'}{$\bm X'$}
\psfrag{e1}{${{\bm e}_{\it 1}}$}
\psfrag{e2}{${{\bm e}_{\it 2}}$}
\rotate[r]{\includegraphics[width=0.5\textwidth]{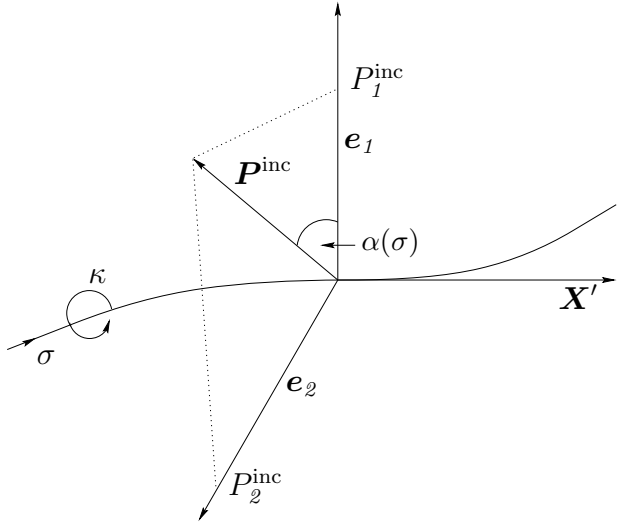}}
\caption{\label{basis} The momentum ${\bm P}^{\rm inc}$ defined from 
the Magnus force action (\ref{Magnus}) is a gauge dependent quantity.
On every point $\sigma$ of the vortex line it can point in an
arbitrary direction of the local co-ordinate plane 
${\bm e}_{\it 1},\, {\bm e}_{\it 2}$.  
This direction is parameterized by the angle $\alpha (\sigma)$.}
\end{figure}    
\end{center}
\subsection{Co-ordinates and momenta}
The rotational part of 
${\bm P}^{\rm inc}$ satisfies  
${\rm rot}\, {\bm P}^{\rm inc}=- m\rho_0\kappa {\bm X}'$ (this follows
from (\ref{bijk})). 
Using a local righthanded co-ordinate basis 
${\bm e}_{\it 1},\, 
{\bm e}_{\it 2},\, {\bm e}_\sigma \equiv {\bm X}'$ with determinant
$g$, 
this yields the gauge invariant relation
\cite{geo}
\begin{equation}\label{p1p2}
\partial_{\it 2} P^{\rm inc}_{\it 1} -\partial_{\it 1} P^{\rm inc}_{\it 2}
= m\rho_0\kappa\sqrt g\,.
\end{equation}
This equation expresses the well-known conjugateness of 
positions and momenta for a vortex line 
in an incompressible fluid (see, {\it e.g.}, \cite{fettquth}).  
It appears here in its generalized form, valid for any choice 
of the co-ordinate basis on the string.
Co-ordinates and momenta in different co-ordinate directions
are no longer independent, just as for the analogous 
electrically charged particle in a very large magnetic field. 
The momentum direction depends on the local co-ordinate basis 
as well as on the particular gauge chosen. 
The Coulomb gauge for $H_V$ in (\ref{H}) is just one possibility.
A convenient and often used gauge  
is to simply set $P^{\rm inc}_{\it 2}=0$ so that 
$P^{\rm inc}_{\it 1}=m\rho_0\kappa \int dX^{\it 2} \!\sqrt g\, ({\bm X})$.
The co-ordinates and momenta can functionally 
depend on each other in a complicated,
non-linear fashion, according to the choice of the co-ordinate system,
which can be local in $\sigma$. 
Of importance in the context treated, though,  
is the fact that all the local momenta 
are given in terms of the local co-ordinates.     

A particularly well-known example \cite{fettquth} for conjugateness of the 
co-ordinates is provided by the Cartesian co-ordinates $X,Y$ of a 
rectilinear line in $z$ direction. 
In the gauge $P_Y=0$, $P_X=m\rho_0\kappa Y= h\rho_0 Y$ and the commutator is 
\begin{equation}
[{\hat X},{\hat Y}]=i(2\pi\rho_0^{(2)})^{-1}\,,
\end{equation}    
where $\rho_0^{(2)}$ is the 2d bulk number density in the $x$-$y$ plane.
For a circular vortex in the gauge $P_R=0$ (cylindrical co-ordinates, 
$\sqrt g ({\bm X})=R$), the canonical momentum per unit length of the 
azimuthal co-ordinate is  
$P_Z(\sigma)=(1/2)m\rho_0\kappa R^2 = (1/2) h \rho_0
R^2$ and we have 
\begin{equation}
[\hat Z, \hat S]= i (2\pi \rho_0^{(3)})^{-1}\,,
\end{equation} 
where $\hat S$ is the operator of the surface area $S=\pi R^2$ 
of the vortex ring. 
\subsection{Volume elements in the tunnelling exponent}
The quantity of relevance for the description of tunnelling 
with constant energy $E$ is the Legendre transform $S_e(E)$
of the Euclidean action $S_e(T)$ ($T$ is the period of motion in
Euclidean time). 
The Euclidean action as a function of imaginary time is   
defined from the real time action by $S_e=-iS[t\rightarrow -it]$. 
\begin{eqnarray}\label{SE}
S_e(E)= S_e(T)-\frac{\partial S_e}{\partial T}\, T= S_e(T) - ET\nonumber\\
=-i \oint\! d\sigma\!\oint d{\bm X}\cdot {\bm P}
=\oint\! d\sigma\!\oint d{\bm K}\cdot {\bm P}\,,
\end{eqnarray}
where we defined the imaginary co-ordinate ${\bm K}=-i {\bm X}$ of the
vortex\footnote{$\bm K$ is not to be confused with a wave vector. We could
have chosen as well to incorporate the $-i$ into the momentum ${\bm P}$.  
Crucial is only that $S_e$ is a real quantity.}.    
The meaning of $S_e(E)$ is most easily understood if we
realize that the 
function $\exp[-S_e(E)/\hbar]$ is an
energy dependent damping factor for a quantum mechanical 
wave function, traveling in imaginary time (with imaginary momentum),
and penetrating a potential barrier (cf., 
in particular, the textbook \cite{ll} for a proper treatment 
of the semiclassical approximation).

The peculiar property of a massless vortex, as seen in the last
subsection, is provided by the fact that all the momenta have to be 
functions of the co-ordinates.   
If we forget about the small correction of the kinetic part to
the momentum and set ${\bm P}={\bm P}^{\rm inc}$, using
(\ref{p1p2}), we can thus bring the Euclideanized 
vortex action in units of Planck's quantum 
of action into the form\footnote{We omit the subscript $V$ for the Euclidean
action.} \cite{geo}:
\begin{equation}
\label{volume}
\frac{S_e (E)}h =
\rho_0\,\int\!\!\!\int\!\!\!\int\! \sqrt g\,
d\sigma\, dZ^1 dZ^2 = \rho_0 \Omega^{(d)}
\end{equation}
The co-ordinate differentials 
are represented by $dZ^1=\cos\alpha \, dK^{\it 1} + \sin\alpha\, dK^{\it 2}$,
$dZ^2=-\sin\alpha\, dX^{\it 1} + \cos\alpha\, dX^{\it 2}$. 
The angle $\alpha (\sigma)$ is the parameter giving the local
direction of the canonical momentum to be chosen  
subject to the constraint (\ref{p1p2}). 
The quantum of circulation was taken to be $\kappa = h/m $. 
More generally,  $\kappa = (N_v/N_s)h/m $, with $N_v$ the vortex 
topological winding number and  $N_s$ the number of
particles with mass $m$ in the elementary boson.

That the only variables 
in the problem are effectively co-ordinates 
yields an expression for the tunnelling exponent which is in
principle very simple. Given
that we can solve for the motion of the local volume element on the line 
$\sqrt g\,d\sigma\, dZ^1 dZ^2 $
along the whole path of the line in co-ordinate space 
as a function of $t,\sigma$,  
and its motion stays completely semiclassical, 
we can calculate the tunnelling exponent 
of the vortex to arbitrary precision. 
The motion of the volume element in (complex) configuration space 
(which represents simultaneously the phase space) determines 
completely the semiclassical tunnelling exponent in the incompressible
limit, that is, if this exponent is dominated by the contributions 
of region I in Fig. \ref{scales}.  
This is always the case if $S_e(E)\gg \hbar$.
We will 
leave this semiclassical domain if 
the dimensionless action 
is approaching some number of order one
or if characteristic variations of the co-ordinates 
take place on scales of the order of the core size $\xi$. 

The integral (\ref{volume}) expresses the  Bohr-Sommerfeld
quantization of the number of particles $N^{(d)}$ contained
in the tunnelling volume $\Omega^{(d)}$:
\begin{equation}\label{N}
S_e(E) = (N^{(d)}+\gamma)h\,\quad
\Leftrightarrow \quad S_e(E)/\hbar = 2\pi (N^{(d)}+\gamma).
\end{equation}
The number $\gamma$ is of the order one and signifies 
the onset of the microscopic quantum regime.
In the semiclassical approximation, $N^{(d)}\gg \gamma $ must hold.   
\subsection{Violation of Galilean Invariance} 
At the absolute zero of temperature, a homogeneous non-relativistic 
superfluid has Galilean invariance, that is, physical contents
are invariant under co-ordinate 
transformations to any reference frame at constant
velocity. If we approach absolute zero, which is  
what is actually realized in experiment, we expect the tunnelling rate  
to make no abrupt change as the temperature is lowered. 
Thus the result for the rate we obtain at $T=0$ 
should also be valid for temperatures slightly above $T=0$. 

Because we can always transform to the
rest frame comoving with the superfluid, 
the tunnelling probability at $T=0$ equals zero if Galilean invariance 
remains unbroken: 
In the rest frame there is a 
tunnelling barrier of infinite height, the logarithmically diverging
vortex self energy. Hence it is necessary to explicitly include the
violation of Galilean invariance by a flow obstacle into 
any calculation of tunnelling rates for Galilean invariant superfluids
at absolute zero. 

The most likely location for the breakdown of invariance to happen 
in a pure superfluid is 
the boundary of the superfluid having some surface roughness
extending on scales much larger than $\xi$. 
We will describe in the section which 
follows the simplest possibility to explicitly solve for the motion
of a vortex in a boundary geometry of varying curvature. This will
provide us with a picture of the geometrical restrictions 
on vortex quantum tunnelling in the semiclassical domain.  
\section{Geometry Dependence of Vortex Quantum tunnelling}\label{geodepend}
It is obvious that to solve in general for the string motion of 
constant energy, according to the equations of motion resulting from 
(\ref{action}) and the Hamiltonian (\ref{H}), 
is a quite complicated task in some nontrivial boundary geometry.
Determining the tunnelling volume $\Omega^{(d)}$ will be
difficult in three spatial dimensions if we do not assume, 
like in \cite{volovik}, the highest possible symmetry of a undeformed,
massless ring vortex in the presence of a sphere.  Even if
we neglect any kinetic and elastic terms of vortex motion, the problem
will require a quite formidable computational effort because to solve 
it, {\it e.g.}, by the image technique  will require a multitude of 
image vortices.     

We will thus cut the problem down on two dimensions 
to display the general properties and dependencies of the tunnelling
volume on external geometry. 
This of course, then, does not describe any possible influence
of waves traveling along the line. Their contribution to the
Euclidean action, however, will be suppressed comparable to 
that of the kinetic term
in (\ref{H}), as long as semiclassicality is retained.   
\subsection{Analytical hydrodynamic solution in two spatial dimensions}
\begin{center}
\begin{figure}[hbt]
\psfrag{sigma}{$\sigma$}
\psfrag{+k}{+$\kappa$}
\psfrag{-k}{-$\kappa$}
\psfrag{Z1}{$Z_1$}
\psfrag{Z1bar}{$\bar Z_1$}
\psfrag{ab2}{$d^2/|Z_1|$}
\psfrag{X}{$X$}
\psfrag{Y}{$Y$}
\psfrag{R}{$d$}
\psfrag{Zplane}{$Z$-plane}
\rotate[r]{\includegraphics[width=0.5\textwidth]{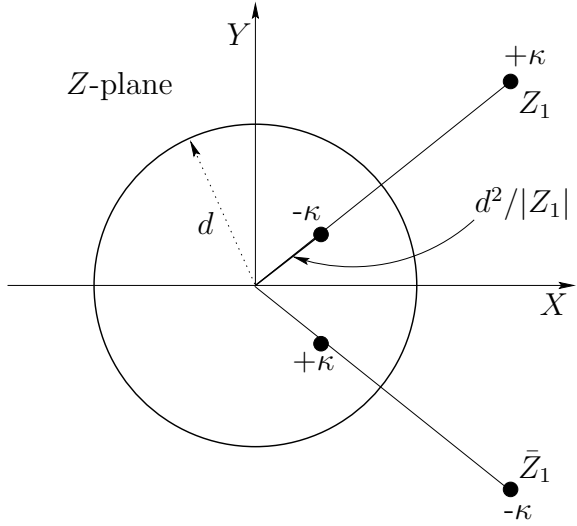}}
\caption{\label{circle} The most simple nontrivial boundary problem solvable 
by the image technique: A point vortex in the half space
$Y>0$, which is filled with liquid, moving  
near a (half-)circle. 
The boundary conditions are satisfied by 
an image vortex at the plane and two image
vortices of opposite strength inside the circle.}
\end{figure}
\end{center} 
The basic solution from which we start is that for a vortex 
in the presence of a half-circle of radius $d$ 
at an otherwise flat boundary (cf. Fig. \ref{circle}\,).
The complex plane of this original solution is called the $Z$-plane. 
The complex potential \cite{milne-t} generated by the image vortices and
acting on the vortex at $Z_1$ is given by 
\begin{equation}
w_i(Z_1)=-i\,\frac\kappa{2\pi}\ln \left[ 
\frac{\left(Z_1-\bar Z_1\right)
\left(d^2/Z_1-\bar Z_1\right)}{{d^2}/Z_1-Z_1}
\right]\,.
\end{equation} 
There a three contributions, coming from the three image vortices
depicted in 
Fig. \ref{circle}. The first factor in the numerator stems from the
 image vortex at the plane $Y=0$ with complex potential $w(Z)=-i(\kappa/2 \pi)
\ln [Z-\bar Z_1]$, the second one is obtained by the circle theorem 
\cite{milne-t} as the image of the original vortex at the circle.
Finally, the potential of the remaining $+\kappa$-circulation vortex
inside the circle, contributing in the denominator of the logarithm, 
completes the image vortex system, again by the 
circle theorem. 
 
The imaginary part of the complex potential 
gives the stream function $\psi = \Im [w]$,  
whereas the real part is the usual velocity potential. The
first term in the denominator of the logarithm is incorporated into the 
static self energy of the vortex,
$E_{\rm self} = (m\rho_0\kappa^2/4\pi)\ln \left(|Z_1-\bar Z_1|/\xi\right)$\,,
which is cut off by $\xi$ and equal to half the energy of a vortex
pair separated by $|Z_1-\bar Z_1|$. The expression for the   
potential in (\ref{H}) is thus $\psi_C 
=-(\kappa /2\pi) \ln\left( \left|(d^2/Z_1-\bar Z_1)/({d^2}/Z_1-Z_1)
\right|\right) $. 

In addition, we superimpose a constant external superflow of velocity
$u$ from right to left. This flow has complex potential  
\begin{equation}
w_u(Z)=u \left(Z+
\frac{d^2}Z\right)\,.
\end{equation} 
From (\ref{H}), we can then infer the energy of the point vortex:
\begin{eqnarray}\label{tildeEZ}
\tilde E(Z_1)= \ln\left| \frac{(Z_1-\bar Z_1)
\left(d^2/Z_1-\bar Z_1\right)}{\xi\left({d^2}/Z_1-Z_1\right)}\right|\nonumber\\
-\frac{4\pi u}\kappa \Im\left(Z_1 + \frac{d^2}{Z_1}\right)\,.
\end{eqnarray} 
The energy is normalized by $m\rho_0\kappa^2/4\pi$, the characteristic
energy of the `particle' vortex 
(corresponding in the electrostatic analogy to the charge squared 
divided by $4\pi$ \cite{myself}).
We omitted in the above equation the contribution of vortex inertia,
$(1/2)\tilde M_0|\dot Z_1|^2$ 
(where $\tilde M_0 = \tilde E_{\rm self}/c_s^2$), 
{\it i.e.} wrote down the energy in the incompressible limit. 

We wish to map by a conformal transformation 
the region outside a boundary 
surface with varying curvature radius, lying in the $z$-plane, 
to the domain outside the circle. Any such transformation can be written
$z= a_0 Z +\sum_{n=0}^\infty b_n Z^{-n}$,  
where $a_0, b_n$ are some coefficients and $Z=d\exp(i\chi)$ is on the
circle \cite{milne-t}. We would like to invert this relation to obtain
the solution for the boundary surface directly from 
that for the circle.
The simplest possibility to do so is supplied by choosing $b_0=0$, $b_n=0$ for
$n>1$ and scaling $a_0$ to unity, 
which leads to the celebrated {\em Joukowski Transformation}
\begin{equation}    \label{Joukowski} 
z=Z-l^2/4Z\,,
\end{equation}
which maps the outside of the ellipse with half axes $a,b$ ($a<b$) 
to the outside of the circle of radius $d=(a+b)/2$. The parameter $l$ is 
defined by $l^2=b^2-a^2$.  
The inversion of the transformation (\ref{Joukowski})  
$2Z= z + \sqrt{z^2+l^2}$ gives, upon insertion 
into the solution for the circle, the expression 
for the vortex energy in the presence of the ellipse. 
It can be seen from the structure of the inverse transformation that 
the use of elliptic co-ordinates, defined by $z=l\sinh\zeta$, 
is most convenient. Here, $\zeta = \chi+i\eta$ is the new complex 
co-ordinate, related to a Cartesian system by  $x=l\sinh\chi
\cos\eta$, $y=l\cosh \chi\sin\eta$. 
The co-ordinate lines are confocal ellipses and hyperbolas,  
and $2Z=l\exp (\zeta)$. 

\begin{center}
\begin{figure}[hbt]
\psfrag{chi}{\normalsize$\chi_1$}
\psfrag{eta}{\normalsize$\eta_1$}
\psfrag{E-E0}{\normalsize$\tilde E-\tilde E_0$}
{\includegraphics[width=\textwidth]{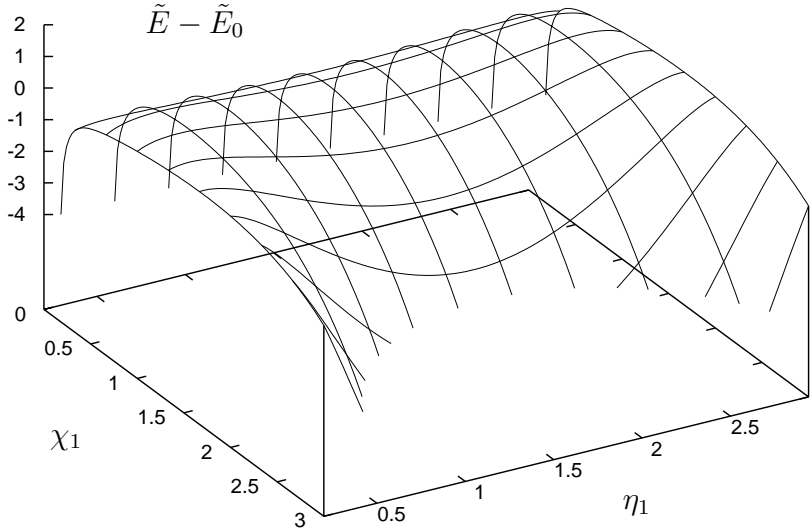}}\vspace*{-1em}
\caption{\label{Escaled} Shape of the potential barrier (\ref{tildeE})
with the choice $\chi_0=0.175$. The corresponding ellipse with 
$b/a\simeq 5.7$ is shown in Fig. \ref{ellipse}. 
The (large) velocity $u=0.08$ is in units of 
$v_L=\kappa/2\pi\xi$ and the ratios $a/\xi=2$, $s/\xi=1$.
The zero of this normalised potential energy is shifted by 
$\tilde E_0$, defined in (\ref{E0}).}
\end{figure}    
\end{center}

In what follows, we consider the solution (\ref{tildeEZ}) in the
half-plane
$y>0$. 
The normalized energy as a function of the elliptic vortex 
co-ordinates $\chi_1,\eta_1$ takes the form
\begin{eqnarray}\vspace*{1em}\label{tildeE}
\tilde{E} (\chi_1,\eta_1) 
 = 
\ln\left[\frac{a+b}{\xi}\,\frac{\exp(\chi_1-\chi_0)|\sin\eta_1| \sinh (\chi_1-\chi_0)}{(\sinh^2(\chi_1-\chi_0)+\sin^2\eta_1)^{1/2}}\right]
\nonumber\hspace*{-4em}\\
-\frac{4\pi u(a+b)}\kappa\sinh
(\chi_1-\chi_0)|\sin\eta_1|\,,
\end{eqnarray}
where $\chi_0 ={\rm artanh} (a/b)$ is the co-ordinate specifying the ellipse
surface. The shape of the energy in the $\chi$-$\eta$ plane is shown in
Fig. \ref{Escaled}. 
The velocity $u$ will henceforth be scaled with the 
characteristic velocity $v_L=\kappa/2\pi\xi$, expressing a {\em velocity
limit} for the validity of the semiclassical approximation. 
(The Landau critical velocity of roton creation $\simeq$ 59 m/s 
at $p\simeq$ 1 bar equals $v_L$ if $\xi\simeq 2.7 $ \AA.)    
The velocity $u=0.08$ chosen for the potential in Fig. \ref{Escaled} 
is large and near critical in the sense that the local 
velocity at the ellipse top is near $v_L$. 
It appears from this graph in real co-ordinate space that the vortex
will predominantly tunnel in the $\chi$-direction 
around the ellipse top ($\eta\approx \pi/2$), 
because there the potential barrier is shallowest.    
\subsection{Geometric restrictions for semiclassicality}
Of course it is possible to invent more intricate conformal
transformations to map the circle solution onto fancy shapes in 
the $z$-plane. But the ellipse already possesses the feature crucial 
for an analysis of vortex motion in a geometry deviating from the highest 
symmetry of Fig. \ref{circle}: It has curvature varying continuously between 
$a/b^2$ in the $x$- and $b/a^2$ in the $y$-direction. 
This leads to a restriction on the possible vortex paths near the 
boundary \cite{geo}, visualized  
in Fig. \ref{ellipse}.
\begin{center}
\begin{figure}[hbt]
{\includegraphics[width=0.75\textwidth]{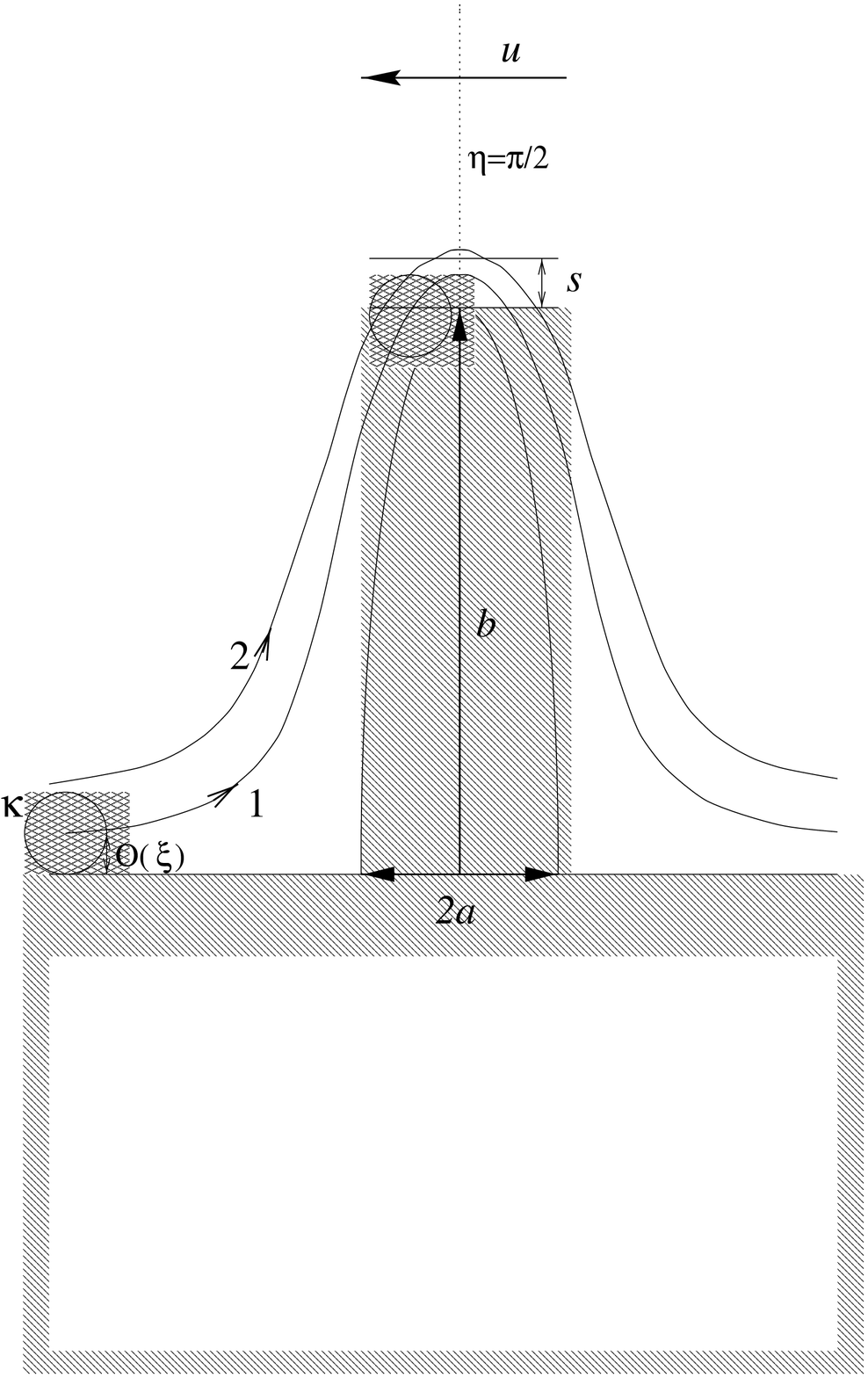}}
\vspace*{-11em}
\caption{\label{ellipse} Two vortex paths of constant energy 
near the ellipse. Whereas
the vortex on path 1 with approximately zero energy, $\tilde E \simeq 0$, 
does not manage to pass by
without coming closer than $\xi$, the second one, having  energy 
$\tilde E= \tilde E_0$, defined in (\ref{E0}), is able to do so.  
The velocity $u$ is to be understood `at infinity'.}
\end{figure} \vspace*{2em}
\end{center}
On the first path we consider           
a vortex having approximately zero energy $\tilde E\simeq 0$, 
starting far away from the ellipse with a distance of the order $\xi$ to the
boundary. As it approaches the
ellipse and tries to get around it, it will inevitably hit the
boundary because it does not have enough energy to cross the sharp
ellipse top at a safe distance. `Safe', that is, such that the
semiclassical approximation can be applied. For that purpose, 
the vortex has to have a distance to the boundary
exceeding the many-body core size $\xi$.
This imprints a restriction on the value of the possible tunnelling
energies for a geometry in which the curvature radius decreases
(the curvature increases) 
while the vortex is passing by:
The energy has to be chosen with a value at least high
enough such that the vortex is able to pass the complete 
boundary at a distance larger than $\xi$. 
Introducing the {\em closest approach distance} $s$ to the boundary 
(see Fig. \ref{ellipse}), and observing that for  
$\delta\chi=\chi_1-\chi_0\ll 1$ the distance of the vortex to the ellipse
top is given by $\delta b \simeq a\delta\chi$, we impose 
$\delta\chi (\eta_1 =  \pi /2)= s/a$, where we assume 
$b\gg a\gg s $ and $s\ge {\rm O}(\xi)$.\footnote{
In the relation $s\ge {\rm O} (\xi)$, the quantity
O($\xi$) means `a value in the order of $\xi$ by definition', 
as there is no sharp distinction between the `inside' and `outside' 
of a quantum vortex.     
Additionally, the choice for the lower limit value of the 
distance $s$ in units of $\xi$ depends on the
value of the core constant $C_0=$ O(1), parameterizing the many-body 
core structure in the vortex energy logarithm 
$\ln[R/(\xi\exp C_0)]$ ($\exp C_0 =1$ for the 
point vortex considered, which has $R = 2Y$).}
Under these conditions, the vortex center on
path 1 in Fig. \ref{ellipse} will pass the top at the distance $\delta b
\simeq (a/b)\xi$, whereas the path 2 vortex with energy 
$\tilde E_0=\tilde E_0 (a,b,s,\xi,u)$,   
\begin{eqnarray}\label{E0}
\tilde E_0
& = &
\ln\left[\frac{a+b}{\xi}\,\exp\delta\chi\, \tanh \delta\chi\right]
-\frac{4\pi u(a+b)}\kappa\sinh\delta \chi
\\
&\simeq & \ln\left[\left(1+\frac{b}{a}\right)\frac s \xi \right]
-\frac{2u}{v_L}\left(1+\frac ba\right)\frac s\xi\,, \nonumber
\end{eqnarray}
passes at $\delta b\simeq s$. The energy $\tilde E_0$ is the energy needed 
by the vortex to remain completely describable in semiclassical 
terms on its way along the ellipse. In general, it depends on $u$,
but in  the low velocity limit $ 2u/v_L\ll 1$, it 
is given by $\tilde E_0\simeq \ln [(b/a) (s/\xi)] $. The effect of this nonzero
energy can be interpreted as the rescaling of $\xi
\rightarrow \xi\exp\tilde E_0$ in (\ref{tildeE}). The energy
is zero {\em with respect to} this rescaled $\xi$. For small $u$, the core
size rescales as $\xi\rightarrow (b/a) s$.

That the validity of the semiclassical approach enforces that we
introduce another geometrical quantity, $s$, is a restriction 
of quite general character.
It is of relevance for any attempt 
to describe tunnelling semiclassically in a realistic, non-spherical 
geometry, {\it i.e.} when the boundary and thus the path of the
tunnelling object near it is not of $S^n$ symmetry. 
A semiclassical description 
is valid only if the quantum core structure of the tunnelling object 
is not touched upon during its motion along the boundary. 
A pinning potential for the vortex 
moving in the superfluid stems in general from some flow obstacle, in our case
the ellipse. 
Any phenomenological {ansatz} for a pinning potential usually employed in 
tunnelling calculations which has curvature 
perpendicular to the applied flow larger than parallel to the flow 
will have to take into account that the object can approach the
surface within its core size, invalidating the collective co-ordinate
description. 

It is to be noted that the energy 
is in units of $m\rho_0\kappa^2/4\pi\simeq 0.82 $ K/\AA\, (at $p\simeq
1$ bar and in three dimensions). For realistic values of the parameters in
(\ref{E0}), the values of $\tilde E_0$ cover the same range as the
phonon-maxon-roton spectrum. Energywise, the trapped vortex thus can not be
distinguished from an elementary excitation. It could have been
excited thermally and remained trapped in a pinning center
during the cool-down of the superfluid to very low temperatures.  
From this and the above analysis  
it follows that it is semantically not appropriate to use the term
`nucleation' if we remain completely semiclassical and 
define it to mean `creation from nothing pre-existent', that is, from
the zero of energy. Experimentally, it will be 
impossible to distinguish the tunnelling 
of a pre-existing small energy vortex at a rough boundary 
from the true nucleation event of a nascent vortex there,  
if no direct means to control the microscopic dynamics can be provided.  

\subsection{The tunnelling area}
\begin{figure}[hbt]
\includegraphics[width=0.75\textwidth]{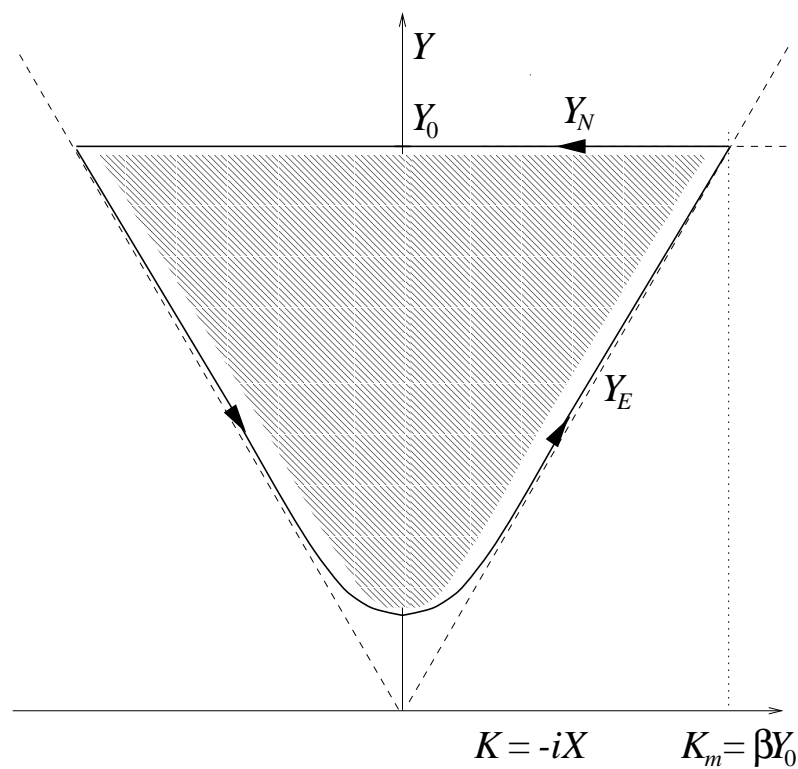}
\caption{\label{area} 
The closed vortex path giving the action (\ref{tunnel}) in the low
velocity limit. The first part $Y_E$ corresponds to the analytically continued
path 2 of Fig. \ref{ellipse} along the ellipse surface. 
The second part of the closed path, $Y_N\simeq Y_0 $, 
obtained from the second solution of (\ref{tildeE})   
for the constant energy in (\ref{E0}), represents the border line to  
a free vortex. 
}  
\end{figure}

In the effectively one-dimensional problem we are considering ($X$ is 
the location of the vortex,  $P_X=m\rho_0\kappa Y= h\rho_0 Y$
its momentum), there always exists a closed vortex path in phase
space. For the tunnelling trajectory this path is in the $K$-$Y$ plane
and the closed vortex path is running underneath the barrier of Fig.
\ref{Escaled}. It has in the limit of small velocities the shape 
shown in Fig. \ref{area}.  

The action (\ref{SE}) in the form of (\ref{volume}),  
which follows from the closed vortex path
in the low velocity limit \cite{geo}, 
\begin{eqnarray}
\frac{S_e}{h}&=&2\rho_0 \int_0^{K_m}\!\!
(Y_{\rm N}-Y_{\rm E})dK\simeq 
\rho_0
\beta  Y_0^2
\nonumber\\
&=&\rho_0 
\left(\frac ab +\frac{s}a \right)
\left(\frac{\kappa}{4\pi u}
\ln\left[\frac{\kappa}{2\pi u}\frac{a/b} {s}
\right]\right)^2\! = \rho_0 \Omega^{(2)}\, , \label{tunnel}
\end{eqnarray}
is bounded from below by the fact that the tunnelling area $\Omega^{(2)}$
can not be arbitrarily reduced without violating semiclassicality, {\it i.e.}
by the introduction of $s$. Geometrically, 
the curvature radius of the vortex path 
near the flow obstacle can not be made arbitrarily small to fit a given 
shape of the asperity on the boundary, so that the eccentricity 
$\epsilon 
\simeq 1-0.5\beta^2$
of the path can not approach unity arbitrarily close. 
The bigger of the two quantities $a/b, s/a\ge \xi/a$ 
decides about the lower limit of $\Omega^{(2)}$,  
attainable in semiclassical tunnelling.

The {\em materialization distance} $Y_0$ of the vortex 
is in the low velocity limit given by (in units convenient for
the discussion of experiments which follows below):
\begin{equation}\label{Y0}
Y_0\simeq\frac {0.8\,\mbox{nm}}{u\, [10\, \mbox{m/sec}]}
\ln \left(\frac{v_L}u\frac{a/b}{s/\xi}\right) \,.
\end{equation}  
In three dimensions, the radius of the materialized vortex 
half-ring, $R_0=(\kappa/2\pi u) \ln (\cdots)
=(1.6\,$nm$/u [10\, \mbox{m/sec}])\ln(\cdots)$, 
bears an additional factor of 2 in front of the logarithm.  
\section{Prefactor estimations; Relation to experimental data}\label{prefactor}
The tunnelling probability has the form 
$P(E)= A \exp [-S_e(E)/\hbar]$, where $A$ is the so-called
prefactor depending on the fluctuations of the quantum variables 
around their classical value. 
Apart from the considerable difficulties in evaluating prefactors
in general, an accurate calculation of $A$ in a dense superfluid like 
helium II is in principle not possible due to the lack of a 
microscopic theory. 
It is, however, feasible to get an idea about the value 
of this prefactor within about two orders of magnitude. This 
is all what we need, because, as we will see below, 
the variations of the tunnelling exponent 
$S_e(E)\gg\hbar$ with the (geometrical) 
parameters of the problem will dominate 
anything which is actually observable (provided of course that a 
semiclassical, large scale description is appropriate).      
The simplest possible idea about the prefactor is gained by
considering the 
frequency $\omega_a$  of a particle
oscillating in a metastable well. Then, neglecting the influence
of dissipation on the vortex motion, within about one order of magnitude 
$A\sim \omega_a$ \cite{schmid}. In the thermal activation limit, {\it i.e.}
in the Arrhenius law case $P=(\omega_a/2\pi)\exp [-U/k_B T]$,
$\omega_a$ is the frequency of oscillations in the metastable well 
against a barrier of height $U$.      
The frequency $\nu_a=\omega_a/2\pi$ can thus be generally understood
as a measure of  
the number of times per second the vortex bounces
against the potential barrier, trying to get free. 
We have no possibility to describe the vortex state (at the boundary) 
quantitatively, but we are able to conclude on the order of magnitude
of $\omega_a$ by taking into account that there exists a surface layer 
of vorticity of width $\xi$. Because the superfluid density goes to
zero at the boundary and heals back within $\xi$, the energy needed
for the activation of vortices vanishes within this distance
\cite{sonin}. The frequency of motion of these 
vortices
should then be of order 
\begin{equation}\label{omegac}
\omega_0
= \frac\kappa{\pi \xi^2}= 4.87 \cdot 10^{11} {\rm sec}^{-1}
\left(\frac{\sigma_{\rm LJ}}\xi\right)^2 \,, 
\end{equation}
which is the cyclotron frequency of vortex motion.  
We scaled $\xi$ with the Lennard-Jones parameter $\sigma_{\rm LJ} 
= 2.556$ \AA\,\,
of the $^4\!$He atomic interaction. Approximately the same estimate 
is obtained if we directly take into account compressibility in the form
of a finite speed of sound. 
The rigidity (spring constant) of the
vortex against deformations will scale as $1/\xi^2$, its mass
as $1/c_s^2$, so that 
\begin{equation}\label{omega0}
\omega_s = \frac{c_s}\xi\simeq 9.23\cdot 10^{11}\mbox{sec}^{-1}
\frac{\sigma_{\rm LJ}}\xi\, , 
\end{equation} 
where we took the pressure to be $p\simeq 1$ bar.
It should be stressed that the approximate coincidence of 
the estimates (\ref{omegac}) and (\ref{omega0}) is a particular 
feature of helium II, where $\kappa \approx 2\xi c_s$. If we adhere 
to the description of a massive vortex of size $\xi$ moving in some
(regularized) potential, the equation (\ref{omega0}) is to be used, 
whereas (\ref{omegac})  is the natural vortex frequency 
associated with the scale $\xi$ alone.

Assuming that the prefactor can vary 
between $A\approx 10^{10}\cdots 10^{12}$ Hz, 
its logarithm  ln$\,A \simeq  23\cdots 28$. 
It is thus obvious that variations of the action in the tunnelling 
exponent with
geometrical factors will outweigh by far any variations in the
prefactor.  
The tunnelling exponent in (\ref{N}) 
involves $N$,  the number of particles in the tunnelling
volume. Corresponding to the conceivable values of the
prefactor, it will have to vary between $N=4\cdots 6 $, say, for tunnelling
events to be observable within a reasonable span of experimental time. Though 
this number is small, it is not impossibly so for the hydrodynamic treatment 
to make sense at least asymptotically ({\em asymptotic} to the
microscopic nucleation regime). 

The experiments in which one hopes to observe intrinsic\footnote{The
term {\em intrinsic} is meant to be the opposite 
of `supported and supplied by pre-existing vortices of macroscopic size',
{\it i.e.} {\em extrinsic}.}
vortex quantum tunnelling at irregular boundaries \cite{ion} 
are using flow orifices of sub-micron size
\cite{exp}. An oscillating superflow of frequency in the order of Hz 
is driven through the orifice. It is observed that at a certain
critical value of the amplitude of the diaphragm driving the flow 
through the hole, there is an instantaneous (on the scale of the
driving frequency) breakdown of the amplitude which is quantized
\cite{AV3}. This quantized dissipation event is associated with 
a vortex generated at the orifice walls, subsequently 
crossing all the streamlines of the flow, thereby causing a phase 
slip event \cite{pwa}, which draws a quantized amount of energy from 
the flow. The critical velocity, at which the
phase slip takes place, is first linearly increasing with decreasing 
temperature and 
then saturates at $T\simeq 150$ mK to a plateau ({\em flat} within
experimental resolution). This plateau has been conventionally 
associated with the quantum tunnelling of half-ring vortices through 
a potential barrier \cite{AIV}. 

In the experiments performed so far, 
the mean critical velocity through the orifice on the plateau 
is of the order 5-10 m/s. 
Though this is not necessarily equal to the velocity $u$ 
at `infinity' (think, {\it e.g.}, of a sharp spike on a smoother
asperity at the orifice wall), the radius $R_0$ 
of the materialized vortex half-ring 
can be estimated to be  
in the order of nanometers. 
This should be sufficiently big for the hydrodynamic 
approach to make sense. The crossover temperature $T_0$ 
from thermal activation to quantum tunnelling is, for negligible
dissipation and small $u$, 
given by  
$2\pi k_B T_0=\hbar \omega_b$, where $\omega_b$ is the oscillation
frequency of vortex motion in the inverted potential (see, {\it e.g.}, 
\cite{schmid},\cite{gorokhovquant}), 
which yields 
$\omega_a \simeq \omega_b = 
8.2\cdot 10^{10} T_0$
[100 mK] Hz.
This is about an order of magnitude smaller than the 
estimates (\ref{omegac}) respectively (\ref{omega0}). There are,
however, 
several facts making a direct comparison of these 
values questionable.     
First of all, these estimates can give only a rough idea about the
true dynamical behaviour of a many-body vortex near the boundary.  
It is conceivable, for example,  that the effective `spring constant'
of the vortex against deformations is lowered compared to the
semiclassical estimate in (\ref{omega0}) because of the many-body 
quantum uncertainty of its position.  
In addition,  we have seen in the preceding section that a  
vortex will in general not be in a state with zero energy at the wall. 
It will rather be in a state with the energy $\tilde E_0$ in 
(\ref{E0}). 
This can also cause a change in the prefactor. 
Furthermore, the prefactor 
is in general a function of driving velocity $u$ and temperature $T$
\cite{gorokhovquant} and at the measured crossover temperature 
in terms of the critical velocity 
not necessarily equal to its value at zero temperature.  
This value will presumably be closer to the 
estimates (\ref{omegac}) and (\ref{omega0}). 

     
\section{Conclusion}
There is no theory of vortex quantum nucleation in the dense Bose
superfluid helium II which could 
resolve the problems connected with theoretical issues   
and experimental facts. There is, though, 
a consistent theory of vortex quantum tunnelling in the semiclassical 
domain, which we represented here in its formal requirements
and geometric implications. It is again to be emphasised that this theory does
not make any assumptions about the dynamical behaviour of the
superfluid on scales approaching the microscopic one. 
This implies on one hand that it necessarily can not supply  a theory of 
vortex nucleation. On the other hand, it gives semiclassical
bounds on tunnelling rates which can be experimentally observed.
It should have become apparent from our discussion that to measure the
consequences of the actual microscopic nucleation event and, 
in particular, influence the (microscopic) 
parameters which govern its probability, 
will be a task quite difficult to realize.
If the tunnelling event takes place on sufficiently large semiclassical 
length and small velocity scales, {\it i.e.} is sufficiently
semiclassical, the only remnant of microscopic dynamics
we can expect to be involved in the tunnelling rate is the core size
$\xi$. But this is only appearing through a logarithm, 
which means that  its influence on the tunnelling rate is quite
subordinate as compared to the tunnelling exponent variation with 
the path geometry of the vortex, imprinted on it  by the flow obstacle. 

The very fact of quantum tunnelling at boundaries 
needs further proof so that the
predictions of tunnelling theory
can be compared to that of classical instability
mechanisms (investigations in this direction are found
under reference \cite{kuznetsov}). 
One such proof could consist in the comparison of critical
velocities for chemically identical orifices of equal global sizes, which
have different microscopic surface structures. 
If the result of such measurements is negative, {\it i.e.} there is no
reproducible difference in critical velocities, there 
is no quantum process taking place 
describable by (semiclassical, hydrodynamic) means
of tunnelling at irregular boundaries.  
The experimental information
at present is too sparse to give a final and conclusive answer.
\begin{ack}
I would like to thank Grisha Volovik and Kolya Kopnin for discussions 
and the Low Temperature Laboratory of the Helsinki University of
Technology, where this work has been done, for hospitality. Financial
support of this research was provided by the Human Capital and
Mobility Programme of the European Union under grant CHGE-CT94-0069.
\end{ack}
\section*{Appendix}
\begin{appendix}
\section{The dual formulation}\label{Adual} 
In this appendix, we introduce the reader to the basics of the dual 
formulation of string dynamics. 
For ease of reading the formulas are given in the conventional index
notation, which will be familiar to the majority of the readership, as
well as in a co-ordinate independent formulation using $p$-forms, 
which shows most clearly their geometrical meaning \cite{MTW}. 
We give the covariant formulae first, 
subsequently reducing them to their Galilean limits, which are
of relevance in superfluids like helium II. 
  
In the hydrodynamic limit, a neutral superfluid 
of spinless bosons 
is described by an order parameter function 
$\phi =\rho^{1/2}(x)\exp [i\theta (x)]$ of the spacetime point $x
=(t,{\vec x})$ (the velocity of light $c\equiv 1$).
Its absolute square is the density $\rho(x)$ and its phase $\theta(x)$ the 
velocity potential of the fluid. The fundamental conservation law
derived from this semiclassical U(1)-description of the superfluid 
is that of the hydrodynamic mass four-current 
${\bm J}\equiv m\rho{\bm v}=\hbar\rho\,{\bm d}\theta =
\hbar \rho\, \partial_\mu \theta\, {\bm d} x^\mu $, where 
$\partial_\mu \equiv \partial/\partial x^\mu$. 
The most familiar mathematical form of mass conservation is provided
by $\partial_t \rho + {\rm div}({\rho \vec v})=0$.   
This is expressed covariantly as   
\begin{equation}
\nabla_\mu J^\mu=0\quad  \Leftrightarrow\quad 
 {\bm d}\wedge\!^*\!{\bm J}=0 
\qquad ({\bm d}\equiv {\bm d}x^\alpha\partial_\alpha)\, .
\label{J}
\end{equation}
That is, the dual of the 
current ${\bm J}=J_\mu {\bm d} x^\mu$ 
is a 3-form $^*\!{\bm J}= 
J^\mu \epsilon_{\mu\nu\alpha\beta}\, {\bm d} x^\nu \wedge {\bm d}
x^\alpha \wedge {\bm d} x^\beta$, 
which is closed. This statement is equivalent to 
a vanishing covariant derivative $\nabla_\mu J^\mu$ of the current vector with 
components $J^\mu$.  
We define the {\em field strength} $\bm H$ by 
$^*\!{\bm J}=m\rho_0\bm H$, where $\rho_0$ is the bulk constant
density. The field $H_{\mu\nu\alpha}$ is totally antisymmetric in its
three indices and has, by definition, only four independent
components. 
In a simply connected region, $\bm H$ is exact, {\it i.e.} it is the exterior 
derivative of a {\em gauge} 2-form $\bm b\, = b_{\mu\nu}\, {\bm d} x^\mu
\wedge {\bm d} x^\nu $: The field strength
$\bm H$ is invariant under gauge transformations ${\bm b}
\rightarrow {\bm b}+{\bm d}\wedge{\bm \Lambda}$, where ${\bm \Lambda}
=\Lambda_\alpha {\bm d} x^\alpha $ is an arbitrary 1-form. 
In components: The replacement $b_{\mu\nu}\rightarrow 
b_{\mu\nu}+\partial_\mu \Lambda_\nu -\partial_\nu\Lambda_\mu $  
leaves $H_{\mu\nu\alpha}$ unchanged. 
We have
\begin{eqnarray}\label{dualtrans}
^*\!{\bm J}= m\rho^*\!{\bm v}=\hbar\rho\, ^*\!{\bm d}\theta
=m\rho_0\,{\bm d}\wedge{\bm b}=m\rho_0 {\bm H},\label{bdef} \\
H_{\alpha\beta\gamma} =\partial_\alpha b_{\beta\gamma}
+ \partial_\gamma b_{\alpha\beta}+ \partial_\beta b_{\gamma\alpha}
=(\rho/{\rho_0})v^\mu\epsilon_{\mu\alpha\beta\gamma}\,,\label{dualtranscomp}
\end{eqnarray} 
which defines the relation of $\rho,\theta$ and $\bm b$.
The dual transformation in the conventional sense \cite{kalb}
is obtained 
if $\rho\rightarrow \rho_0$, {\it i.e.} if we consider compressibility
to be negligible in the domain of interest. 
Then, ${\bm b}$ has only one degree of freedom
corresponding to the order parameter phase $\theta$. For our purposes,
this reduction to a single degree of freedom 
amounts to not considering the cores of vortices. This is reasonable,
because in the dense superfluid helium II the semiclassical
U(1) description is not valid in the vortex core.  

The superfluid we are dealing with is only Galilei invariant, not
Lorentz invariant, so that by writing (\ref{dualtranscomp}) 
with $\rho=\rho_0$ for spatial and temporal
indices separately and taking the Gali\-lean limit,   
we get the relations
\begin{eqnarray}\label{bijk}
-\sqrt g\, n_{ijk} = 
\partial_k b_{ij}+\partial_j b_{ki}+\partial_i b_{jk}=H_{ijk}\,, \\
\label{stream}
\sqrt g \, n_{ijk}\, v^i =  \partial_k\psi_j -\partial_j \psi_k\,,
\end{eqnarray}
where $\psi_i \equiv b_{0i}$ is the vectorial version of the stream
function of classical hydrodynamics \cite{milne-t}. The
determinant of the spatial 
co-ordinate system we are using is designated $g$ and $n_{ijk}=
n_{[ijk]}=\pm 1$ is the unit antisymmetric symbol.
\end{appendix}

\end{document}